\documentclass[prl,twocolumn,showpacs,preprintnumbers]{revtex4}
\usepackage{graphicx}% Include figure files
\begin{document}
\preprint{Proc.\ TH-2002, Paris, July 2002, to be published in
Ann.\ Henri Poincar\'e}
\title{Applications of Ideas from Random Matrix Theory \\ to Step Distributions on
``Misoriented" Surfaces}
\vspace{-0.3cm}
\author{T.~L.\ Einstein}
\email[E-mail: ]{einstein@umd.edu}
\homepage{http://www2.physics.umd.edu/~einstein}
\affiliation{Department of Physics, University of Maryland,
College Park, MD 20742-4111 USA\\}
%\vspace{-0.3cm}

\date{\today}
%\vspace{-1.5cm}
\begin{abstract}

    Arising as a fluctuation phenomenon, the equilibrium distribution of
meandering steps with mean separation $\langle \ell \rangle$ on a
``tilted" surface can be fruitfully analyzed using results from
RMT.  The set of step configurations  in 2D can be mapped onto
the world lines of spinless fermions in 1+1D using the
Calogero-Sutherland model.  The strength of the
(``instantaneous", inverse-square) elastic repulsion between
steps, in dimensionless form, is $\beta(\beta-2)/4$.  The
distribution of spacings $s\langle \ell \rangle$ between
neighboring steps (analogous to the normalized spacings of energy
levels) is well described by a {\it ``generalized" Wigner
surmise}: $p_{\beta}(0,s) \approx a s^{\beta}\exp(-b s^2)$. The
value of $\beta$ is taken to best fit the data; typically $2 \le
\beta \le 10$. The procedure is superior to conventional Gaussian
and mean-field approaches, and progress is being made on formal
justification.  Furthermore, the theoretically simpler step-step
distribution function can be measured and analyzed based on exact
results.  Formal results and applications to experiments on
metals and semiconductors are summarized, along with open
questions. (conference abstract)

%*Contact info: e-mail: einstein@umd.edu
%                     FAX: 1-301-314-9465

\end{abstract}
%\pacs{PACS Number(s): 05.40.+j,61.16.Ch,68.35.Md,68.35.Bs}
\maketitle

%%%%%%%%%%%%%%%%%%%%%%%%%%%%%%%%%%%%%%%%%%%%%%%%%%%%%%%%%%%%%%%%%%%%%%%%

\section{Introductory Synopsis}
\label{sec-introsyn}

For over a decade we have studied the correlation functions of
steps on vicinal surfaces of crystals, that is to say crystals
that are intentionally misoriented from high-symmetry directions.
Of special experimental interest is the distribution of
separations of adjacent steps, corresponding to $p_{\beta}(0,s)$
of random matrix theory \cite{MehtaRanMat}. Traditionally, surface
scientists have described it by a Gaussian, with a controversy
arising about the relationship between the width of the Gaussian
and the strength of the repulsion between steps. This controversy
is resolved, and a better representation of the step distribution
obtained, by characterizing the problem with the
Calogero-Sutherland models and using the simple expression of the
Wigner surmise, with [seemingly for the first time] {\it general}
values of $\beta$ in the range of 2 to 10. This range is rarely
considered by random-matrix theorists. The step-step correlation
function resulting from the Calogero-Sutherland models, for which
exact but numerically intractable solutions exist, can also be
applied to data.

\section{Background}
\label{sec-background}

On vicinal crystals there are a sequence of terraces oriented in
the high-symmetry direction and separated by steps typically one
atomic layer high. If the spacing between adjacent steps (the
``terrace width") is denoted $\ell$, then the mean spacing
$\langle \ell \rangle$ is 1/$\tan \phi$, where $\phi$ is the
angle by which the surface is tilted from the high-symmetry
direction.  For theoretical modeling it is convenient to think of
the crystal as being a simple cubic lattice.  (The extension to
realistic symmetries is not difficult but muddies the
discussion.)  By [``Maryland"] convention,  the steps run on
average in the $\hat{y}$ direction, so that the step spacings
$\ell$ are in the $\hat{x}$ direction.  (The high-symmetry
terrace plane is then $\hat{z}$.)  For metals it is a decent
approximation to take the total energy of the system to be
proportional to the number of nearest neighbor bonds.  (For
semiconductors corner energies can play a role, but these are
insignificant for the properties discussed in this paper.)  Then
the ground-state configuration is a set of perfectly straight
steps (uniformly spaced because of repulsions we shall discuss
shortly).  At low temperatures, the predominant excitation are
kinks in the steps, each costing energy $\epsilon$.  At higher
temperatures, atoms and vacancies appear on the terraces (each
with energy 4$\epsilon$ in the simple model), but these are
neglected here.  The resulting model is called the
terrace-step-kink model (TSK) or terrace-ledge-kink model in the
literature.

\begin{figure}
\includegraphics[scale=0.4,angle=90]{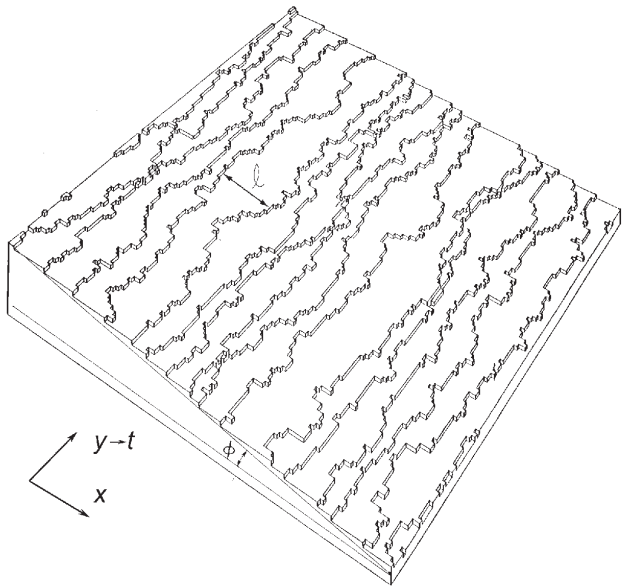}
\caption{\label{fig:TSK} Illustration of a vicinal generated by a
Monte Carlo simulation of the terrace-step-kink (TSK) model,
showing that the elementary thermal excitation is the kink, a
unit excursion of a step in the $x$ direction. (Adapted from
Fig.\ 1 of Ref.\ 27) For experimental STM (scanning tunneling
microscopy) images of a vicinal surface; see, e.g., Fig.\ 1 of
Ref.\ 12, Fig.\ 1 of Ref.\ 15, and Fig.\ 1 of Ref.\ 34. In the
Gruber-Mullins approximation, only one step meanders, the others
being straight along the $y$ direction and uniformly spaced. (Cf.
Fig.\ 9 of Ref.\ 12.) In the fermion picture, the $y$ direction
is viewed as time-like. In the step continuum model, the steps
meander continuously rather than by discrete deviations (i.e.,
the kinks are smoothed).}
\end{figure}

In this low-temperature picture, the number of steps is fixed (by
the miscut angle in an experiment and by screw periodic boundary
conditions in a numerical simulation).  Thus, the steps never
start or stop, and they all have the same orientation (say up):
there are no anti-steps (or down-steps).  It is convenient and
fruitful to imagine the $\hat{y}$ direction as time-like and to
view the configurations of steps as the worldlines of a
collection of a single kind of particle in one spatial dimension
(viz.\ $\hat{x}$) evolving in time.  For physical reasons, the
steps cannot cross each other (since that would involve atoms
suspended above the terrace held with just a couple lateral
bonds).  Assuming also that the steps cannot merge to form
double-height steps, we can view the evolving 1D particles as
spinless fermions. (Equivalently, in 1D, they can be treated as
hard bosons, a viewpoint exploited in a study of step bunching
\cite{szs}. Moreover, the system is highly reminiscent of soliton
lines in 2D incommensurate crystals \cite{pt}.)

The obvious next question is how these spinless fermions
interact. A typical way to characterize the distribution of steps
is by their terrace-width distribution (TWD), i.e. the
probability of finding the next step a distance $\ell$ away.  This
probability distribution is denoted $P(s)$, where $s \equiv
\ell/\langle \ell \rangle$ denotes the step spacing normalized by
the [only] characteristic length in the $\hat{x}$ direction.  By
construction, $P(s)$ is normalized and has unit mean. (It
ultimately corresponds to $p(0,s)$ of random-matrix theory.) For
the ground state, $P(s)$ is essentially a delta function at
$s$=1. If the steps are imagined as straight (``uncooked
spaghetti") and deposited randomly with probability 1/$\langle
\ell \rangle$, then $P(s)$ becomes exp(-$s$).  In fact, however,
at finite temperature the steps do meander, leading to well-known
entropic (or steric) repulsion due to the underweighting in the
partition function of steps on the verge of crossing.  This
effect is particularly noticeable as $s$ approaches 0: $P(s)$ now
vanishes with power-law behavior rather than growing to unity.
For $s \gg 1$ $P(s)$ must decay faster than exp(-$s$) to preserve
the unit mean; analogies with random walkers suggest exp(-$bs^2$)
as the form of the tail \cite{MEF}. There are many arguments to
show that the entropic repulsion per length is proportional to
$(k_BT)^2/\tilde{\beta}\ell^2$; here the step stiffness
$\tilde{\beta} \equiv \beta(\theta) +d^2\beta/d\theta^2$, with
[orientation-dependent] free energy per length conventionally
called $\beta(\theta)$ in surface studies.

In addition to the entropic repulsion, there generally is an
elastic interaction, also having the inverse-square form
$A/\ell^2$, due to repulsion between force dipoles intrinsic to
the steps \cite{Rickman,mp}: Viewed from a continuum picture, the
step involves a pair of sharp angles which could lower their
energy by healing to form more obtuse angles.  This process leads
to a strain field around the step in which atoms in the terraces
bounding the step (both above and below) tend to move away from
the step.  Atoms between two steps are then frustrated since they
are pushed in opposite directions by the pair of steps.  Thus,
they cannot relax fully, leading effectively to a repulsion, with
$A$ proportional to the product of the step dipoles.

In the fermion picture, the strength of the elastic repulsion
only enters in the dimensionless combination

\begin{equation}
\tilde{A} \equiv \tilde{\beta}A/(k_BT)^2.
\end{equation}

\noindent The elastic and entropic repulsions do not simply add
together:  As the strength $A$ of the elastic interaction
increases, the chance of steps coming near each other decreases,
thereby reducing the entropic repulsion. Hence, the dimensionless
strength of the inverse square repulsion in essence
\cite{ftntpsq6} becomes \cite{WmsErr,CL}
\begin{equation}
\label{eq:all} \tilde{A} \rightarrow \frac{1}{4}\left[1+\sqrt{1+4\tilde{A}}\right]^2 \, .
\end{equation}

In situations where the surface has a metallic surface state [one
which crosses the Fermi energy], there can also be an oscillatory
[in sign] interaction between steps, albeit with a 1/$\ell^2$
envelope \cite{RZ92,TLE-U}.  Such an interaction destroys many of
the scaling properties, particularly that the form of $P(s)$ is
well defined, i.e.\ that plots of $\langle \ell
\rangle\hat{P}(\ell)$ vs.\ $\ell/\langle \ell \rangle$ are
independent of $\langle \ell \rangle$.  Further discussion of
this digression is beyond the scope of this paper.

Knowledge of the elastic strength $A$ is crucial to adequate
characterization of a stepped surface.  It is one of just three
parameters (the stiffness $\tilde{\beta}$ being another) of the
widely-applied step continuum model \cite{JW99} and underlies the
``2D pressure", which determines surface morphology (e.g.,
whether steps ``bunch") and drives kinetic evolution.  It is
generally believed that the repulsions can be adequately
approximated as acting between pairs of steps at the same value
of $y$, i.e. instantaneously in the 1D fermion representation.
This approximation is clearly crucial for viability of the
fermion approach.

The simplest way to estimate $P(s)$ was developed over a third of
a century ago. In treating polymers in 2D, deGennes pointed that
the energy of a step (or a polymer) could be described as the
line integral over path-length ingrements times $\beta(\theta)$.
Expanding to lowest order, we get a constant (the straight step)
plus what amounts to a kinetic energy proportional to
$\tilde{\beta} \, (dx/dy)^2$, the mass-like stiffness times the
1D velocity squared.

In the mean-field (i.e. single active step with all others fixed
at spacings $\langle \ell \rangle$) approximation
\cite{Gruber67}, Gruber and Mullins (GM) reduced the ``free
fermion" case ($A$=0) to the familiar elementary quantum
mechanics problem of solving the Schr\"odinger equation for a
spinless fermion of mass $\propto \tilde{\beta}$ in a hard-wall
box of length 2$\langle \ell \rangle$.  The probability density
of the ground state (the squared ground-state wavefunction) is
just $P_{GM-0}(s) = \sin^2(\pi s/2)$, for $0 \le s \le 2$;
remarkably, the associated ground-state energy is precisely the
entropic repulsion.  For $\tilde{A}$ above about 3/2, the problem
becomes that of a particle in a parabolic potential, i.e. a
simple harmonic oscillator, giving $P_{GM} \propto
\exp[-(s-1)^2/2w_{GM}^2]$, with $w_{GM} \propto
\tilde{A}^{-1/4}\langle \ell \rangle$ \cite{Bartelt90}.  These
parametric results were dramatically verified for Si(111): not
only was the TWD measured at one misorientation well fit by a
Gaussian, but the TWD at a second misorientation was well
described by another Gaussian whose width was simply rescaled by
the change in $\langle \ell \rangle$, with $no$ refitting
\cite{WangWms,Metois}.

However, the Grenoble group pointed out a flaw in Gruber-Mullins
approach in the limit of very strong $\tilde{A}$ \cite{IMP98}.
There the entropic repulsion becomes negligible and the steps can
be imagined as meandering independently (albeit by a small
amount).  Then the variance of the TWD should not be just
$w_{GM}^2$, but twice that amount, since both steps bounding a
terrace are meandering.  (The actual increase is less than
two-fold due to lingering anticorrelations.) Hence, a fit of the
TWD width in the Gruber-Mullins approximation would underestimate
$\tilde{A}$ by a factor of over 3 in the limit of very large
$\tilde{A}$.  Based on roughening theory, the Saclay group
\cite{Barbier} concluded that the underestimate, for more general
$\tilde{A}$, was somewhat smaller (about 2).

Meanwhile, Ibach noted that the TWD of the free-fermion
distribution could be far better approximated by an expression
$\propto s^2 \exp(-bs^2)$, in essence the Wigner surmise for free
fermions, than by $P_{GM-0}(s) = \sin^2(\pi s/2)$.  How this
comes about becomes evident by recasting the preceding model in
terms of the celebrated \cite{CGbook} Calogero-Sutherland models
\cite{C69,Suth71Cal,Suth71Suth}.

\section{Connection to Random Matrix Theory via Calogero-Sutherland Models}

In his study of the spacings of energy levels in nuclei, Wigner's
starting inspiration was to consider ensembles of dynamical
systems governed by different Hamiltonians with common symmetry
properties \cite{Guhr98}.  The three generic Hamiltonian
symmetries are orthogonal, unitary, and symplectic.  The key
ingredient is level repulsion: two states connected by a
non-vanishing matrix elements repel each other by an amount
determined by the symmetry of the Hamiltonian.  While inapplicable
to average quantities, the approach is appropriate for the
fluctuation of a large number of energy levels. These
fluctuations should become independent of the specifics of the
level spectrum and the weight factors, and so should exhibit a
universal form depending only on symmetry.  This idea can also be
derived from maximum-entropy arguments \cite{Guhr98}.

One can draw a correspondence between the fluctuations of energy
spacings between adjacent energy levels in nuclei and the
fluctuations of spatial separations between adjacent fermions in
1D systems.  (In other chaotic systems, as well, there is a
correspondence between energy and spatial spacings.) The
Calogero-Sutherland model describes spinless fermions in 1D which
interact via an inverse square potential.  Specifically,
Sutherland's version \cite{Suth71Cal} of the Calogero Hamiltonian
\cite{C69} for the fluctuations of uniformly-spaced fermions on an
infinite line is:

\begin{eqnarray}
{\cal H}_{\rm CS} = -\sum_{j=1}^N \frac{\partial^2}{\partial x_j^2} &+& 2
\frac{\beta}{2}\left(\frac{\beta}{2}-1\right)\sum_{1 \le i<j \le N}
 (x_j - x_i)^{-2} \nonumber \\
 &+& \omega^2 \sum_{j=1}^N x_j^2,
\label{eq:HCS}
\end{eqnarray}

\noindent  in the limits $N\rightarrow \infty$ and $\omega
\rightarrow 0$. (In Calogero's original model \cite{C69}, the
last term ($\propto \omega^2$) is summed over particle
separations $(x_j-x_i)^2$ rather than deviations from integer
positions $x_j^2$.)  The ground-state wavefunction for this
Hamiltonian is

\begin{equation}
\Psi_0 = \prod_{1 \le i<j \le N}\left| x_j - x_i \right|^{\beta/2}
\exp \left(-\frac{1}{2}\omega \sum_{k=1}^N x_k^2 \right)
\end{equation}

\noindent The ground-state density $\Psi_0^2$ is recognized as a
joint probability distribution function from the theory of random
matrices for Dyson's {\it Gaussian} ensembles \cite{Dyson69}.

The Sutherland Hamiltonian\cite{Suth71Suth} similarly describes
spinless fermions on a circle of radius $L$ (with $L\rightarrow
\infty$), having inverse-square interactions along chords:

\begin{equation}
\label{eq:HS} {\cal H}_{\rm S} = -\sum_{j=1}^N
\frac{\partial^2}{\partial x_j^2} + 2
\frac{\beta}{2}\left(\frac{\beta}{2}-1\right)\frac{\pi^2}{L^2}\sum_{i<j}\left[
\sin \frac{\pi (x_j - x_i)}{L}\right]^{-2}
\end{equation}

\noindent In this case the ground-state wavefunction has the
Jastrow form

\begin{equation}
\Psi_0 = \prod_{i<j}\left| \sin \frac{\pi (x_j -
x_i)}{L}\right|^{\beta/2}, \quad x_j > x_i
\end{equation}

\noindent With the definition $\theta_i \equiv 2\pi x_i/L$, the
ground-state density $\Psi_0^2$ can be written as
\begin{equation}
\Psi_0^2 = \prod_{i<j}\left|e^{i\theta_j} - e^{i\theta_i}
\right|^{\beta}
\end{equation}

\noindent This ground-state density is again a joint probability
distribution function from the theory of random matrices, now for
Dyson's {\it circular} ensembles \cite{Dyson62}.

 From Eqs.\ (\ref{eq:HCS}) and (\ref{eq:HS}), the dimensionless
interaction $\tilde{A}$ can be identified as
$(\beta/2)(\beta/2-1)$. Conversely, $\beta$ is just the quantity
inside the brackets in expression (\ref{eq:all}). These models
can be solved exactly {\it only} for the special cases $\beta$ =
1, 2, or 4, corresponding to orthogonal, unitary, or symplectic
symmetry of the ensemble.  In the literature of random matrix
theory\cite{MehtaRanMat}, the TWD $P(s)$ is often denoted
$p(0,s)$, the probability of finding two levels separated by $s$
with zero intervening levels. Note also that $\beta$ was used in
the previous section to denote the step free energy per length.
(Presumably the letter was chosen to emphasis the similarity to
but difference from the surface free energy per area $\gamma$.)
Dyson \cite{Dyson62,MehtaRanMat} chose $\beta$ to denote this
exponent to indicate the analogy to an inverse temperature (of a
Coulomb gas on a circle).  While we follow this convention here
to minimize the chance of confusion, in our applications to
stepped surfaces the exponent is called $\varrho$ to minimize
confusion for that audience.  (The stiffness $\tilde{\beta}$ is
essentially unrelated to the exponent $\beta$!)

Random matrix theory leads to exact solutions for the ground state
properties of the three special cases, in particular for $P(s)$.
There are even prescriptions that allow one to generate numerical
representations of arbitrary accuracy \cite{Dyson70,Joos91}.
However, these expressions involve terms in a series and are not
useful for fitting experimental data.  Wigner surmised that
$P(s)$ has the simple form

\begin{equation}
P_{\beta}(s) = a_{\beta} s^{\beta}e^{-b_{\beta} s^2} \label{eq:WS}
\end{equation}

\noindent for the special cases $\beta$= 1, 2, and 4.  The
constants $a_\beta$ associated with normalization of $P(s)$ and
$b_\beta$ producing unit mean are:
\begin{equation}
     a_\beta = \frac{2\left[\Gamma
\left(\frac{\beta +2}{2}\right)\right]^{\beta +1} }
                   { \left[\Gamma
\left(\frac{\beta +1}{2}\right)\right]^{\beta +2} }
\quad\mbox{and}\quad
     b_\beta =
   \left[\frac{\Gamma \left(\frac{\beta +2}{2}\right)}
              {\Gamma \left(\frac{\beta +1}{2}\right)}\right]^2  \, .
\label{e:abr}
\end{equation}

\noindent From Eq.\ (\ref{eq:WS}) one can readily find analytic
expressions for the moments and other measurable properties of
$P(s)$ \cite{EP99}.

The argument for Eq.\ (\ref{eq:WS}) focusses on the Jacobean
associated with a change of variables for the Gaussian-ensemble
probability distribution function $p({\cal H}) \sim \exp [-b N\,
{\rm tr}({\cal H}^2)]$ from the eigenenergies to the combinations
of the matrix elements.  For an orthogonal ensemble ($\beta$=1)
with $N$=2, the integrand has a Dirac delta function with
argument $s -[(h_{11}-h_{22})^2 +4h_{12}^2]^{1/2}$, which
vanishes for $s$=0 only when the two [squared] independent
variables do.  Hence, $P(s) \propto s$, corresponding to a
circular shell in parameter space and leading to $\beta$=1.  For
unitary ensembles there is an additional independent parameter
since $h_{12}^2$ becomes $(\Re {\rm e}\, h_{12})^2 + (\Im {\rm
m}\, h_{12})^2$. Hence, $P(s) \propto s^2$, corresponding to a
spherical shell in parameter space, i.e.\ $\beta$=2. Exact for
$N$=2, these arguments are still fine guides for large $N$
\cite{Zwer98}: As seen most clearly by explicit plots as in
Haake's text \cite{Haake91}, $P_1(s)$, $P_2(s)$, and $P_4(s)$ are
excellent approximations of the exact results for orthogonal,
unitary, and symplectic ensembles, respectively, and these simple
expressions are routinely used when confronting experimental data
\cite{Guhr98,Haake91}.  (The agreement is particularly impressive
for $P_2(s)$ and $P_4(s)$, which are germane to stepped
surfaces.) Furthermore, Eq.\ (\ref{eq:WS}) has a similar variance
at very large $\beta$ to that predicted by the Grenoble group,
and more generally approaches the form of a Gaussian for $\beta$
not too small\cite{ERCP01}.

\begin{figure}
\includegraphics[scale=0.6]{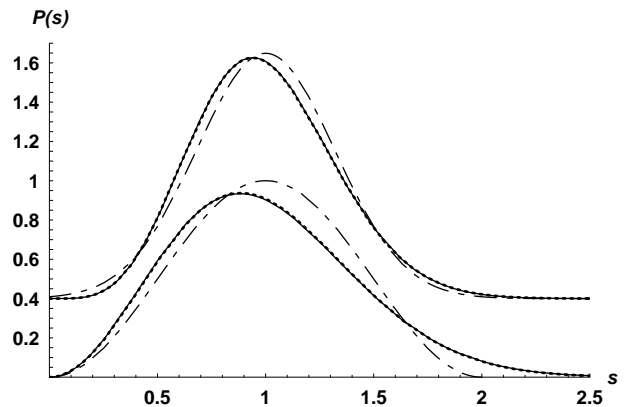}
\caption{\label{fig:Ps} $P(s)$ vs.\ $s \equiv \ell/\langle \ell
\rangle$ for the exact ``free-fermion", $\tilde{A}$=0 result
(solid curve), the Gruber-Mullins approximation $\sin^2(\pi
s/2)$(long-short dashed curve), and the $\beta$=2 Wigner surmise
(Eq.\ (8)) result (dotted curve), barely distinguishable from the
exact result). Offset upward by 0.4 for clarity, an analogous
plot of the exact result for $\tilde{A}$=2, the Gruber-Mullins
Gaussian approximation $(24/\pi^2)^{1/4}\exp(-\surd 24 (s-1)^2)$
(long-short dashed curve), and the $\beta$=4 Wigner surmise
result (dotted curve).  (Adapted from Fig.\ 1 of Ref.\ 31).}
\end{figure}

The essential innovation in our work is the conjecture that the
generalized Wigner surmise expressed in  Eq.\ (\ref{eq:WS}) for
{\it arbitrary} values of $\beta$ provides a good approximation
for these non-special values.  For these general values, there is
no recourse to formal justification from symmetry arguments.
(Indeed, it remains mysterious whether there is some deep
underlying reflection of the random-matrix symmetry and the
corresponding Calogero-Sutherland models at the special values of
$\beta$.)  The only way to test our hypothesis is to generate
numerical data.  This task has occupied us for some time.  We
have used both Monte Carlo and transfer matrix computations of TSK
models \cite{ERCP01,HCRE}.  The details of these studies are
beyond the scope of this paper, but the main result is that the
generalized Wigner surmise does better than any of the earlier
Gaussian approximations in accounting for the dependence of the
TWD's variance---the quantity typically measured by
experimentalists---on the value of $\tilde{A}$.  In particular,
the numerical data confirms that the proportionality coefficient
between the width $w$ of the TWD and $\tilde{A}^{-1/4}$ (or more
accurately, between $w$ and $\beta^{-1/2})$ is {\it not} constant
but increases slowly with $\tilde{A}$.

Obtaining the experimentally measurable variance $\sigma^2$ of
$P(s)$---essentially $w^2$ when the Gaussian approximation is
viable---analytically from Eq.\ (\ref{eq:WS}), we\cite{RCEG} can
use a series expansion to derive an excellent estimate of
$\tilde{A}$ from $\sigma^2$:
\begin{equation}
   \label{eq:as}
   \tilde{A}
     \approx  \frac{ 1}{16} \left[(\sigma^2)^{-2}
           - 7 (\sigma^2)^{-1}
           + \frac{27}{4}
           + \frac{35}{6} \sigma^2\right]  \, ,
\end{equation}
\noindent with all four terms needed to provide a good
approximation over the full physical range of $\tilde{A}$, from
near zero up to around 16, corresponding to $\beta$ ranging from
2 up to about 8.  The Gaussian methods described earlier
essentially use just the first term of Eq.\ (\ref{eq:as}) and
adjust the prefactor.

We have successfully applied these ideas (both direct fits to Eq.\
(\ref{eq:WS}) and use of Eq.\ (\ref{eq:as})) to experimental data
by several different groups \cite{ERCP01}, focussing primarily on
extensive results for vicinal Cu (001) and (111) obtained by
Giesen, Ibach, and collaborators at FZ-J\"ulich
\cite{RCEG,Giesen00}.

\section{Pair Correlation Functions}

While the TWD is usually the easiest quantity to measure, it is
generally not the easiest to calculate since it is essentially a
many-body correlation function due to the demand that there be no
fermion between the two fermions separated by $\ell$ (or $s$). The
pair correlation function $h(s)$ (also called $1-Y(s)$ in
random-matrix literature\cite{MehtaRanMat}) indicates the
probability of finding a fermion a normalized distance $s$ from a
reference fermion, regardless of whether there are any between
them.  For a ``perfect staircase" this would just be a set of
delta functions at positive integers.  More generally, $h(s)$
increases like $s^{\beta}$ for $s \ll 1$ just like $P(s)$.  There
are then a sequence of peaks at integer values of $s$ with dips
between them.  The amplitude of the crests and troughs relative to
the mean density of fermions decreases as $s$ increases.  For most
purposes, but not here, it suffices to use the harmonic
approximation \cite{KO82,SB99,Forres92} (which is essentially
similar to the Grenoble approximation):

\begin{eqnarray}
\label{e:harm} \langle \ell \rangle h_{\beta}(s) &=& \sum_{m \ne
0}(B_m\pi)^{1/2} \exp
[-B_m (s-m)^2]; \\
\nonumber
%B_m = \frac{(\pi/2)^2 \beta}{\gamma +\log (2\pi m) -{\rm ci}(2\pi m)}
\frac{\pi^2 \beta}{4 B_m}=\gamma\! &+&\!\log (2\pi m)\! -\!{\rm
ci}(2\pi m) =\sum_{j=1}^{\infty}(-1)^{j+1}\frac{(2\pi
m)^{2j}}{2j(2j)!},
\end{eqnarray}

\noindent where $\gamma \simeq 0.577$ is Euler's constant and ci
is the cosine integral.  Note that $B_m$ is proportional to
$\beta$, so that peaks become sharper and higher with increasing
repulsion.  Also, $B_m$  decreases slowly but monotonically with
increasing $m$.  While useful for many applications
\cite{KO82,SB99,Forres92}, this approximation for $h_{\beta}(s)$
proved inadequate for deducing $\beta$ from data.

Forrester \cite{Forres92} found an exact solution for even values
of $\beta$.  This expression, which involves Selberg correlation
integrals, takes nearly a page of text, so is not reproduced
here.  We do note that the ratio of the exact prefactor of
$s^{\beta}$ in $h_{\beta}(s)$ to $a_{\beta}$ in $P_{\beta}(s)$ is
close to unity, lending further support to the utility of the
generalized Wigner surmise expression \cite{ftntharatio}.

Subsequently, Ha and Haldane \cite{HH,h94,Les} derived an exact
solution for any rational value of $\beta$.  In principle, this
result also contains information about the ``dynamic"
correlations between fermions (i.e.\ between steps at different
values of $y$). Unfortunately, their expression is even longer
than that of Forrester, and so also is not reproduced here.
Moreover, it seems to be numerically intractable, so of little
``practical" use.

Fortunately, Gangardt and Kamenev \cite{GK01} recently derived
the following asymptotic expansion for $h(s)$:

\begin{eqnarray}
\langle \ell \rangle h_{\beta}(s) &\sim& -\frac{1}{\pi^2\beta s^2}
+2\sum_{j=1}^{\infty}\frac{d_j^2(\beta)}{(2\pi s)^{4 j^2/\beta}}
\cos(2 \pi
j s); \\
\nonumber d_j(\beta) =& \Gamma& \! \! \! \! \! \! \left(\! 1 +
\frac{2j}{\beta}\right)
\prod_{m=1}^{j-1}\left(\frac{2m}{\pi\beta}\right) \sin \!
\left(\frac{2\pi m}{\beta}\right) \Gamma^2 \! \! \left(\frac{2
m}{\beta}\right). \label{eq:GK}
\end{eqnarray}

\noindent This formula turns out, remarkably, to be useful well
below the asymptotic limit, and provides a good
approximation---much better than the harmonic approximation---for
$s>1/2$; it can be patched onto $a_{\beta}s^{\beta}$ for  $s \le
1/2$.

In confronting data with Eq.\ (\ref{eq:GK}), one must beware the
limited number of steps in experimental images.  In our
application \cite{CohenSchr}, there were only about a half dozen
steps per image.  Hence, we included an {\it ad hoc} linear decay
envelope to account for this limit on $h(s)$ at large $s$.  The
system in question involved steps on Si(111) at very high
temperatures, necessitating a clever device to compensate for
evaporation.  For a variety of reasons, the difficulty of the
measurement led to data that presented exceptional challenges for
analysis.  One problem was that steps sometimes disappeared from
the image, leading to the concern that the TWD might appear wider
than it in fact was (since sometimes second-neighbor spacings
rather than adjacent ones might be counted).   The investigation
showed that electromigration from the current which heated the
sample tended to keep the steps apart, leading to correlation
functions indicative of a larger [effective] $\tilde{A}$ than
predicted from extrapolation of results at lower temperature.

\section{Concluding Comments}

We have seen that steps on misoriented surfaces can be
represented by fluctuating spinless fermions in 1+1D.  Thus,
step-step correlation functions can be expressed in terms of the
Calogero-Sutherland model and thereby related to random matrix
theory.  In particular, the terrace width distribution,
identified as a joint probability distribution, can be well
described by the generalized Wigner surmise, evidently not only
for the special cases related to orthogonal, unitary and
symplectic ensembles but for arbitrary fermion repulsion
strength.  While perhaps reminiscent of some interpolation
schemes, such a generalization has not, to my knowledge, been
made before.  Furthermore, most of these interpolation schemes
have involved systems with mixed symmetries (typically orthogonal
and unitary), with values of $\beta$ between 1 and 2 (and also
between 0 and 1, where $\beta$=0 indicates Poissonian [random
``uncooked spaghetti"] behavior). For stepped surfaces the
physically interesting range {\it starts} near $\beta$=2 and goes
past 4 up to 8 or even 10.  This range has received minimal (if
any) theoretical attention.

While the discovery of exact solutions is always intellectually
exciting and captivating, they are often of little use in
confronting experiments---physical or numerical---unless the
formulas are numerically tractable.  Thus, even when a problem is
formally solved, neither theoreticians or funding agencies should
view it as completed until the formalism can be rendered in a way
that allows the extraction of numbers, even if approximately.

A major open question is the ``deeper meaning" of the generalized
Wigner surmise of Eq.\ (\ref{eq:WS}), since there is no evident
symmetry to argue for the simple expression.  Howard Richards has
taken the lead in trying to derive this result from an effective
Hamiltonian \cite{beyond}.  This approach should lead to the
development of the form of $P(s)$ for systems with higher-order
terms in the repulsion potential and even potentials with an
oscillatory term such as when metallic surface states mediate the
interaction.  (In systems in which ``non-instantaneous"
interactions (in directions other than $\hat{x}$ are significant,
the usefulness of the mapping to 1D fermions is likely to break
down.)  It will be interesting to see whether expressions similar
to the generalized Wigner surmise arise in other areas of physics.
For example, while considering the probability distribution of
stock market returns in the Heston model with stochastic
volatility (in econophysics research), Yakovenko showed that the
probability distribution of the variance---which is derived from
a Fokker-Planck equation, has the form of the generalized Wigner
surmise \cite{DY02}! (In this case the value of $\beta$ was 1.6,
so in a different regime from stepped surfaces.

While Eq.\ (\ref{eq:WS}) assumes that $s$ is continuous,
experiments and numerical simulations involve discrete lattices.
We have shown \cite{ERCP01,HCRE,RCEG} that so long as $\langle
\ell \rangle$ is at least 4 lattice spacings, there are no
significant complications due to discreteness and that assuming
continuous $s$ does not distort the analysis.  In particular, the
roughening transition to a facet that occur for discrete lattices
\cite{Barbier,VGL} does not happen in the range of physical
parameters \cite{HCRE}.

Another skirted issue concerns how many steps interact. While the
entropic interaction {\it ipso facto} involves just adjacent
steps, the Calogero-Sutherland models
\cite{C69,Suth71Cal,Suth71Suth} and the Saclay approach
\cite{Barbier} assume energetic interactions between all steps.
The Gruber-Mullins \cite{Gruber67,Bartelt90} and Grenoble
\cite{IMP98,EP99} approximations allow all steps or just adjacent
ones to interact.  Most Monte Carlo and transfer matrix
simulations assume that just adjacent steps interact.  In the
Gruber-Mullins approximation, the curvature of the potential
involves the inverse 4th power of the distance from neighboring
steps, so that the effective strength of the repulsion increases
by $\zeta(4) = \pi^4/90 \approx 1.08$ when interactions between
all steps rather than just adjacent ones are included
\cite{ERCP01}; in the Grenoble approximation it increases by about
1.10.  These changes are relatively small, but the good agreement
is curious between the variance predicted by Eq.\ (\ref{eq:WS})
and our Monte Carlo results with just adjacent steps interacting
\cite{HCRE}.

For surface scientists a fundamental objective is to show
consistency between the value of $A$ deduced from the TWD and
that predicted from surface stress using Marchenko's formula
\cite{mp}.  However, most experimental systems do no possess the
elastic isotropy assumed by that classic result.  Hence, this
goal has been elusive.

Another interesting problem is what happens when the mean
direction of the steps (i.e.\ $\hat{y}$) does not correspond to a
high-symmetry direction of the surface.  We have recently found
that the stiffness computed from standard near-neighbor bond
lattice models (Ising or SOS models) underestimates the
experimentally observed stiffness by a factor of about 4 for the
square-net face of copper \cite{DGIE}.  The most likely
explanations are significant long-range interactions or perhaps
local relaxations leading to 3-atom effects.  Furthermore, it is
by no means obvious that the assumption of only ``instantaneous"
interactions between steps is justifiable for substantial
azimuthal misorientation.

\section*{Acknowledgment}
This conference paper is based on work done in collaboration with
O.\ Pierre-Louis, H.L.\ Richards, Hailu Gebremariam, S.D.\ Cohen,
M.\ Giesen, H.\ Ibach, J.-J.\ M\'etois, R.D.\ Schroll, N.C.\
Bartelt, B.\ Jo\'{o}s, and E.D.\ Williams, and supported by
NSF-MRSEC Grant DMR 00-80008. Some of the work of HG, SDC, and
TLE has also been partially supported by NSF Grant EEC-0085604. I
am grateful for enlightening conversations with many colleagues,
most notably M.E.\ Fisher, S. Bahcall, H.\ van Beijeren, and V.\
Yakovenko, and, at TH-2002, M.L.\ Mehta, C.A.\ Tracy, and I.\
Smolyarenko. I thank A.\ Kaminski for helpful comments on the
manuscript.

%%%%%%%%%%%%%%%%%%%%%%% Bibliography %%%%%%%%%%%%%%%%%%%%%%%%%%%%%%%


\begin{thebibliography}{10}
\bibitem{ftntpsq6} There is an additional overall factor of $\pi^2/6$, so that the combined
repulsion is $(\pi k_BT)^2/(24\tilde{\beta}\ell^2)[1+(1+4\tilde{A})^{1/2}]^2$
\cite{WmsErr,CL}.
\bibitem{ftntharatio} The ratio is a slowly decreasing function of
$\beta$, around 1.04 at $\beta$=1, crossing 1 around $\beta$=4,
and around 0.97 around $\beta$=10.
\end{thebibliography}
\end{document}